\documentclass[useAMS,usegraphicx,usenatbib]{mn2e}
\bibpunct{(}{)}{;}{a}{}{,}
\usepackage{fontenc}
\usepackage{graphicx}
\usepackage{lscape}
\usepackage{longtable}
\usepackage{rotating}
\usepackage{colordvi}
\usepackage{color}
\usepackage[english]{babel}
\usepackage{amsmath}
\usepackage{amssymb}
\usepackage{amsfonts}
\usepackage{epsfig}
\usepackage{subfig}
\usepackage{natbib}

\def \apjl {ApJL}
\def \apss {ApJ Suppl. Ser.}

\def \zmax {z_{\rm max}}

\title[Solutions of the Lane-Emden equation]{Numerical solutions of the modified Lane-Emden equation in $f(R)$-gravity}

\author[R.\ Farinelli et al.]{R.  Farinelli$^{1,2}$, M. De Laurentis$^{3,4}$, S. Capozziello$^{3,4}$, S. D. Odintsov$^{5,6,7}$\\
$^{1}$ISDC Data Center for Astrophysics, Universit\'e de Gen\`eve, chemin d'\'Ecogia 16, 1290 Versoix, Switzerland\\
  $^{2}$Dipartimento di Fisica e Scienze della Terra, Universit\'a di Ferrara, Via Saragat 1, I-44122 Ferrara, Italy \\
$^{3}$Dipartimento di Fisica, Universit\'a di Napoli ``Federico II'', \\
Compl. Univ. di Monte S. Angelo, Edificio G, Via Cinthia, I-80126 Napoli, Italy\\
$^{4}$INFN Sez. di Napoli, Compl. Univ. di Monte S. Angelo, Edificio G, Via Cinthia, I-80126 Napoli, Italy,\\
$^5$Instituci\'o Catalana de Recerca i Estudis Avan\c{c}ats (ICREA), Barcelona, Spain\\
$^6$Institut de Ci\`encies de l'Espai (CSIC-IEEC), Campus UAB,\\ Facultat de Ci\`encies, Torre C5-Par-2a pl, E-08193 Bellaterra (Barcelona), Spain\\
$^7$Tomsk State Pedagogical University (TSPU), Tomsk, Russia.}

\begin{document}
\date{}

\maketitle

\label{firstpage}

\begin{abstract}
The modified Lane-Emden equation for stellar hydrostatic equilibrium in $f(R)$-gravity is numerically solved by using an iterative procedure. Such an integro-differential equation  can be obtained in the weak field limit approximation of $f(R)$-gravity by considering a suitable polytropic equation of state in the modified Poisson equation.
The  approach allows, in principle, to deal with   still unexplored  self-graviting systems that could account for exotic stellar structures that escape standard stellar theory.
\noindent

\end{abstract}

\begin{keywords}
equation of state -- gravitation -- hydrodynamics -- methods: numerical
\end{keywords}

\section{Introduction}

 Extended Theories of Gravity (ETGs) are acquiring more and more interest
 in modern astrophysics and cosmology due to several inescapable issues and shortcomings that, in principle, could be addressed by revising or extending General Relativity (GR) \citep{poisson,axially,quadrupolo,BH,ivan,SCF,enzo,CARDO,greci}.
In particular,  observational and theoretical issues related to dark matter (DM) and dark energy (DE) could be encompassed in a new picture where gravitational interaction is depending on the scale of the self-gravitating systems without requiring further material ingredients \citep{annalen} that, up to now, seem extremely elusive (see the recent results of the LUX Collaboration 2013). In particular, one of the most famous observational issues which needs  DM, in addition to the visible one, concerns dynamics of stellar objects in galaxies, a problem arising since the beginning \citep{oort32,Oort}.
The disagreement between the observed and expected mass needed to explain dynamics soon
revealed to be present even at cosmological scales \citep{zwicky33}.
Besides, the observed accelerating behavior of the Hubble flow  \citep{Riess} needs to be addressed at fundamental level either 
by revising the Standard Cosmological Model or finding out cosmic fluids capable of giving rise to cosmic acceleration.

It is however worth pointing out that the possibility to consider modifications of the Newtonian
dynamics should be taken into account not only when considering the coupling of matter and
gravity field of  each-other influencing distinct objects, but also in the case of self-gravitating systems.
These can be of different types and sizes: collapsing molecular clouds, instability
strip and main sequence stars, hyper-massive stars, and post-evolved very compact objects such
as magnetars \citep{jeans,ivan2}. In particular, exotic compact objects are not only matter of speculative
theories but are gaining observational evidence of their presence \citep{muno06} and very often escape 
interpretation in the framework of standard Newtonian theory.
Specifically, the structure and evolution of very massive stars is a key field of investigation
for interpreting peculiar objects like Gamma Ray Burst (GRB). The nature of progenitors could play a crucial role in determining the energetics and hydrodynamical processes related to the central
engine in order to provide the observed enormous amount of $10^{51\div54}$ erg of  electromagnetic
energies.
Suitable candidates for GRB progenitors are thought to be Wolf-Rayet stars with hydrogen-stripped envelopes 
and masses of the helium core $M \sim 15\div30 M_{\odot}$. 

Additionally,   numerical simulations 
claim for the possibility of pair-instability supernovae with $M \ga 130M_{\odot}$ as possible progenitor,
if not for all, at least for a part of GRBs \citep{cct10}.
It is thus evident that addressing the problem of stellar stability over a wide range of
masses and radii has important implications for several  aspects of their pre- and
post- evolutionary history.
These theoretical issues can be considered valid both at  descriptive and predictive
level:  from one side there is the need to explain the existence of objects which seem to have 
masses higher than their  expected value ({\it i.e.} Volkoff magnetars);
from the other side,  providing theoretically stable stellar models (not yet observed) can have important
implications  both for  stellar evolution and for testing gravity.
In particular,  modified gravity theories may lead to extra branch of solutions for
 neutron stars as recently pointed out in \cite {Astashenok}.
 
In general, the stability of stellar structures is strictly related to  the hydrostatic equilibrium
conditions. Equations governing  such equilibrium contain  the gradient of the gravitational
potentials and are, ultimately, related to the field equations governing the coupling
between matter and gravitational field.
In the Newtonian theory, if the thermodynamical relations of stellar structures are determined by a polytropic equation
of state between pressure and density, the gravitational potential $\Phi$ is determined
through the well known Lane-Emden equation (LEE).
The weak-field approximation in GR  case leads to a generally non-linear second-order differential equation
governing the behaviour of the gravitational potential over the polytropic stellar structure \citep{kippe}.

However, when the Hilbert-Einstein action is modified considering a function  $f(R)$ of the Ricci scalar, in particular, if we expand this function up to $R^2$, the weak field
equations lead to a system of coupled modified Poisson equations between $\Phi$ and the scalar
curvature invariant $R$ \citep{poisson}. This means, equivalently,  that two gravitational potentials have to be considered being the field equations of fourth order \citep{PRnostro}.
This system gives rise to a \emph{modified Lane-Emdem equation} (MLEE)
for $\Phi$ which contains an additional convolutional term related to the Yukawa-like correction in
the gravitational potential \citep{annalen}. 
The importance of this term depends on the interaction length-scale over which it acts.

In this paper, we will consider numerical solutions of MLEE coming from $f(R)$-gravity.  We want to show that, beside standard stellar structures, coming from the Newtonian limit of GR, other structures could be addressed. The eventual detection of such structures could constitute a formidable test for alternative theories of gravity, in particular for ETGs.

The paper is organized as follows: in Section \ref{due} we
review briefly how the MLEE  comes out in the context of $f(R)$-gravity.
In Section \ref{sect_algorithm},  a description of the algorithm used to numerically integrate the MLEE is given. 
In Section \ref{results} we show the numerical results  for some values of polytropic index  comparing them with GR results. In particular, we discuss the role of the Yukawa-like correction coming out in the Newtonian potential.
Stellar masses and gravitational energy binding are calculated in Section \ref{cinque}. Finally, discussions
and conclusions are drawn in Section \ref{sei}.

\section{The modified Lane-Emden equation}
\label{due}

The modified Lane-Emden equation (MLEE) has been  derived in the context  of  $f(R)$-gravity, for self-gravitating
systems in the linear approximation of weak field \citep{poisson}. 
We refer the reader to that paper for the mathematical details and its derivation. 
Here,  we briefly report the the most relevant steps. \\
\noindent In the metric approach, the $f(R)$ field equations are
obtained by varying the action with respect to
$g_{\mu\nu}$ \citep{PRnostro,reviewodi,reviewodi1,reviewmauro,reviewvalerio}. We get

\begin{eqnarray}\label{fieldequationHOG}
\left\{\begin{array}{ll}
f'(R)R_{\mu\nu}-\frac{f(R)}{2}g_{\mu\nu}-f(R)_{;\mu\nu}+g_{\mu\nu}\Box f'(R)=\mathcal{X}\,T_{\mu\nu}^{(m)},\\
\\
3\Box
f'(R)+f'(R)R-2f(R)\,=\,\mathcal{X}\,T^{(m)}\,,
\end{array}\right.
\end{eqnarray}
where the second equation is the trace.
Here,
$T_{\mu\nu}=\frac{-2}{\sqrt{-g}}\frac{\delta(\sqrt{-g}\mathcal{L}_m)}{\delta
g^{\mu\nu}}$ is the the energy-momentum tensor of matter, while
$T=T^{\sigma}_{\,\,\,\,\,\sigma}$ is the trace,
$f'=\frac{df(R)}{dR}$, $\Box={{}_{;\sigma}}^{;\sigma}$ and
$\mathcal{X}\,=\,8\pi G$\footnote{Here we use the convention
$c\,=\,1$.}. The conventions for the Ricci tensor is
$R_{\mu\nu}={R^\sigma}_{\mu\sigma\nu}$ while for the Riemann
tensor is
${R^\alpha}_{\beta\mu\nu}=\Gamma^\alpha_{\beta\nu,\mu}+...$. The
affinities are the standard Christoffel symbols
$\Gamma^\mu_{\alpha\beta}=\frac{1}{2}g^{\mu\sigma}(g_{\alpha\sigma,\beta}+g_{\beta\sigma,\alpha}
-g_{\alpha\beta,\sigma})$. The adopted signature is $(+---)$ \citep[see][]{landau}.

The field equations (\ref{fieldequationHOG}) can be rewritten as

\begin{eqnarray}
\label{ET}
G_{\mu\nu}=\mathcal{X} \left(T_{\mu\nu}^{(m)}+T_{\mu\nu}^{(c)}\right)\,,
\end{eqnarray}
where $G_{\mu\nu}$ is the Einstein tensor and  $T^{ (c)}_{\mu\nu}$ is the curvature energy-momentum tensor that comes out when we introduce curvature higher order terms in the Lagrangian \citep{CS}. It is defined as 
\begin{eqnarray}
&&\mathcal{X}T^{(c)}_{\mu\nu}=\left[1-f'(R)\right]R_{\mu\nu}+\frac{1}{2}g_{\mu\nu}\left[f(R)-R\right]+\nonumber\\&&+\nabla_\mu \nabla_\nu f'(R)-g_{\mu\nu}\Box f'(R)\,,
\end{eqnarray}
and  vanishes as soon as $f(R)=R$.
As we can see, in the right side of  equation (\ref{ET}), there are two fluids: a standard matter fluid and a curvature fluid. In this way we can treat even the ETGs as the GR in the presence of two sources \citep[see][for details]{gae}.
Note that these quantities satisfy the Bianchi identities, in fact,
 using the following  properties
 \begin{equation}[\nabla_\gamma, \nabla_\beta]V^\alpha = - R^\alpha_{\,\,\, \rho \beta\gamma} V^\rho\,, \end{equation}
  where $V^\alpha$ is a generic vector, and
 \begin{equation}[\nabla_\mu, \nabla_\nu]f'(R)=0\,, \end{equation}
  it is straightforward to show that

 \begin{equation}\label{divTc}
    \nabla^\mu (\mathcal{X} T^{(c)}_{ \mu\nu})=0\,,
 \end{equation}

\noindent
so that the total energy-momentum tensor can be written as
  \begin{equation}
 T_{\mu\nu}=T^{(m)}_{\mu\nu}+T^{(c)}_{\mu\nu}\,,
  \end{equation}
and then $\nabla^\mu  T_{\mu\nu}=0$.
Let us now  perturb the metric tensor up to $c^{-2}$ so that the Ricci scalar becomes
  \begin{equation}
  R\sim R^{(2)}+{\cal O}(4)\,,
  \end{equation}
 and the $n$-th derivative of the Ricci function can be developed as
\begin{eqnarray}
f^{n}(R)\,&\sim&\,f^{n}(R^{(2)}+\mathcal{O}(4))\sim\,\nonumber\\
&\sim&\,f^{n}(0)+f^{n+1}(0)R^{(2)}+\mathcal{O}(4)\,,\nonumber\\
\end{eqnarray}
where $R^{(n)}$ indicates a quantity of order $\mathcal{O}(n)$.
From lowest order of field equations (\ref{fieldequationHOG}),  we
have $f(0)\,=\,0$ which trivially follows from the assumption that the space-time
is asymptotically Minkowskian. Equations (\ref{fieldequationHOG})  at $\mathcal{O}(2)$ - order (Newtonian level)
become

\begin{eqnarray}\label{PPN-field-equation-general-theory-fR-O2}
\left\{\begin{array}{ll}
R^{(2)}_{tt}-\frac{R^{(2)}}{2}-f''(0)\triangle
R^{(2)}=\mathcal{X}\,T^{(0)}_{tt}\,,\\\\
-3f''(0)\triangle
R^{(2)}-R^{(2)}=\mathcal{X}\,T^{(0)}\,,
\end{array}\right.
\end{eqnarray}
where $\triangle$ is the Laplacian in the flat space, $R^{(2)}_{tt}=-\triangle\Phi$ and for simplicity we set $f'(0)=1$. We recall that the energy momentum for a perfect fluid considered is the following
\begin{equation}
T_{\mu\nu}=(\epsilon+p)u_{\mu} u_{\nu}-pg_{\mu\nu}\,,
\end{equation}
where $p$ is the pressure  and $\epsilon$ is the energy density of perfect fluid. Then we have

\begin{equation}\label{HOEQ}
\left\{\begin{array}{ll}\triangle\Phi+\frac{R^{(2)}}{2}+f''(0)\triangle R^{(2)}\,= \,-{\cal X}\rho\,,\\ \\
3f''(0)\triangle R^{(2)}+R^{(2)}\,=\,-{\cal X}\rho\,,\end{array}\right.
\end{equation}
where $\rho$ is the mass density. We note that $f''(0)\,=\,0$ we have the Newtonian mechanics: $\triangle\Phi\,=\,-4\pi G\rho$. 
This equation can be considered the modified Poisson equation for $f(R)$-gravity.

\noindent From the Bianchi identity (satisfied for first line of eq. \ref{fieldequationHOG}) we find

\begin{equation}\label{equidr}
\frac{\partial p}{\partial x^k}=-\frac{1}{2}(p+\epsilon)\frac{\partial \ln g_{tt}}{\partial x^k}\,.
\end{equation}

Let us suppose now that  matter satisfies a polytropic equation $p\,=\,K\,\rho^\gamma$ with $K$ free independent parameter or a constant with fixed value and $\gamma$ the polytropic exponent. Since $\epsilon\,=\,\rho c^2$,  equation (\ref{equidr}) satisfies this relation
\begin{equation}
\frac{\gamma K}{\gamma-1}\rho^{\gamma-1}\,=\,\Phi\,\,\,\,\rightarrow\,\,\,\,\rho=\left[\frac{\gamma-1}{\gamma K}\right]^{\frac{1}{\gamma-1}}\Phi^{\frac{1}{\gamma-1}}= A_n\Phi^n
\label{densita}
\end{equation}
where $n$ is the polytropic index, which is defined by $n=\frac{1}{\gamma-1}$. We remember that in non relativistic limit we have $\gamma=\frac{5}{3}$ and $n=\frac{3}{2}$ while for the relativistic limit $\gamma=\frac{4}{3}$ and $n=3$. Note that for this cases, where the equation of state a polytropic form, the polytropic constant $K$ is fixed and can be calculate from a natural constants.\\
Here, using the above equation (\ref{HOEQ}) and after some calculations, we get the general form of the MLEE which reads as
\begin{eqnarray}
&& \frac{d^2w(z)}{dz^2}+\frac{2}{z}\frac{dw(z)}{dz}+w(z)^n= \label{le_equation} \nonumber\\&&
=\frac{m\xi_0}{8~z}\int_0^{\xi/\xi_0} dz' z' \{ e^{-m\xi_0 | z-z'|}-e^{-m\xi_0 | z+z'|}\} w(z')^n\,,\nonumber\\
\label{LE}
\end{eqnarray}

\noindent
where $z=R/\xi_0$  is a dimensionless radius given in units of the
polytropic radius, which is defined as

\begin{equation}
 \xi_0=\sqrt{\frac{3}{16 \pi G}} K^{1/2} (1+n)^{1/2} {\rho_c}^{\frac{1-n}{2n}}.
\label{poly_radius}
\end{equation}

\noindent
The dimensionless function $w(z)=\Phi/\Phi_c$   gives the gravitational potential normalized
to its central value

\begin{equation}
 \Phi_c={\rho_c}^{1/n} (n+1) K,
\end{equation}
where ${\rho_c}$ is the central density, while $K$ and $n$ are the constant and the index
of the polytropic equation of state

\begin{equation}
 P=K \rho^{\gamma}~~~~~~~~~~~~~~~ \gamma=1+\frac{1}{n}.
\end{equation}

The quantity $m \xi_0$ in the right-hand side of equation (\ref{le_equation}) is
 related to the effective length scale of the Yukawa-like correction term
to the Newtonian potential. Defining $L_{\rm Y} = 1/m$, we may write $m \xi_0= \xi_0/L_{\rm Y}$;
when $L_{\rm Y}/\xi_0 \gg 1$,  $m \xi_0 \rightarrow 0$, the gravitational potential 
asymptotically becomes Newtonian, and  the right-hand side
term of equation (\ref{le_equation}) goes to zero, in turn recovering
the GR case.

\noindent It is well know that analytical solutions for the classical Lane-Emden equation
exist only for three values of the polytropic index $n$, equal to 0, 1 and
5, respectively. Of these, only $n=0$ and $n=1$ give finite-radius solutions.

\noindent The situation of course gets even more complicated in the case of  MLEE,
for its full integro-differential nature-- in this case analytical solutions
are available only for $n$=0 \citep[see][]{poisson}.

\noindent A systematic investigation of polytropic stellar structures thus requires the developement
of numerical methods aimed at obtaining general solution of equation (\ref{le_equation}). This is actually
necessary for both $f(R)$ and GR models, given that, {\it e.g.}, degenerate non-relativistic ($n=3/2$)
and relativistic ($n=3$) polytropic equations of state are needed to  build up  stellar
models.
\section{Description of the algorithm}
\label{sect_algorithm}

Let us numerically solve  equation (\ref{le_equation}) using an iterative
procedure described as follows.
If we define $w_i(z)$ the function at the $i^{\rm th}$-iteration and
define the integral term of the right hand side as $\aleph [w(z')]$,
the iteration algorithm can be written as

\begin{equation}
 \frac{d^2w_i}{dz^2}+\frac{2}{z}\frac{dw_i}{dz}+w_i^n=\frac{m\xi_0}{8~z}\aleph [w_{i-1}].
\label{iter_equation}
\end{equation}

In the framework of a numerical procedure, the term on the right-hand side of
 equation (\ref{iter_equation}) can be viewed as a source term of a  
non linear second order differential equation.
We can now rewrite the equation as a system of two first-order differential equations in the form

\begin{equation}\label{system} 
\large{\left\{\begin{array}{ll}\frac{dw_i}{dz}=k_i,\\ \\
\frac{dk_i}{dz}=-\frac{2}{z}k_i - w_i^n + \frac{m\xi_0}{8~z} \aleph [w_{i-1}],
\end{array}\right.}
\end{equation}


\noindent
with boundary conditions $w_i(0)=1$ and $k_i(0)=0$, respectively.\\
The last term on the right-hand side of the second equation is proportional to 
the  quantity  $\aleph [w_{i-1}]$ which is determined by
the function $w_{i-1}$ obtained at the $(i-1)^{\rm th}$-iteration.
At the first iteration $i=1$, an initial guess for $w_0$ must be given, and
we choosed $w_0=1$.
The solution of the system (\ref{system}) is obtained via a coupled  $4^{\rm th}$-order
Runge-Kutta integration method up to the value $z_{\rm max}=\xi/\xi_0$ with
the above mentioned initial conditions

The regular and smooth behaviour of the function $w(z)$ in its domain of definition allows to compute
the integral on right-hand side of equation (\ref{le_equation})  using the Gauss-Legendre
quadrature according to the general rule

\begin{equation}
 \int_0^{\zmax} f(z) dz \approx  \frac{\xi}{2 \xi_0} \sum\limits_{j=1}^n q(z_j) f( \frac{\xi}{2 \xi_0} z_j + \frac{\xi}{2 \xi_0} ),
\end{equation}

\noindent
where 

\begin{equation}
 q(z_j)=\frac{2}{1-{z_j}^2 [{P_n}'(z_j)]^2},
\end{equation}

\noindent
are the weights and $z_j$ is the $i^{\rm th}$ root of the Legendre polynomial $P_n(z)$.
The values of the function $f(z)$ (see eq. \ref{le_equation}) at the nodes $z_j$
are computed by mean of a linear interpolation.

\noindent

\begin{figure}
\hspace{-0.4cm}
\centerline{\includegraphics[width=6.5cm, angle=-90]{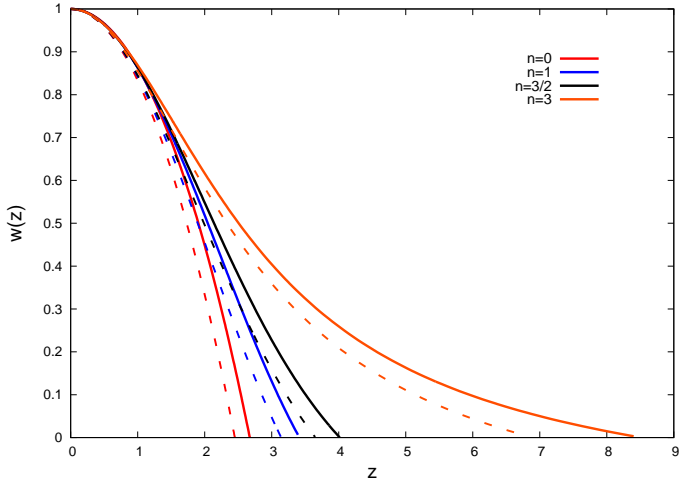}}
\caption{Numerical solutions of the Lane-Emden equation (\ref{le_equation}) for different values of the polytropic index $n$.
For each value of $n$, two solutions are reported: $m \xi_0=1$ (continuum line) and $m \xi_0=0$ (dashed-line). 
The first case gives w(z) corresponding to field equations with $f(R)$, the second to $f(R)=R$.}
\label{wfunct_solutions} 
\end{figure}

\begin{figure}
\hspace{-0.4cm}
\centerline{\includegraphics[width=6.5cm, angle=-90]{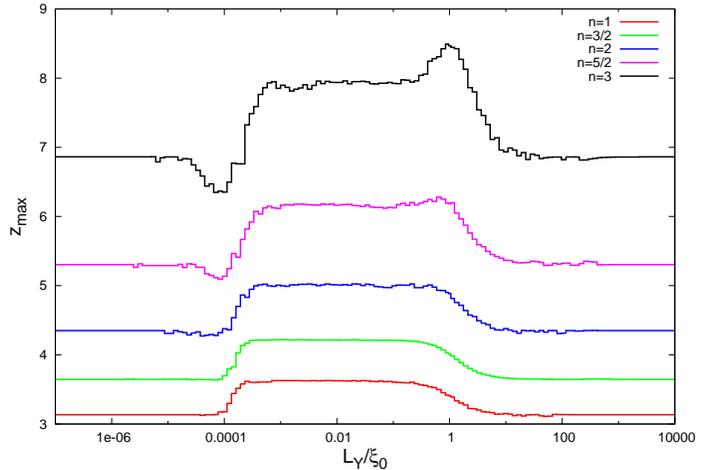}}
\caption{Stellar radius in units of the polytropic radius $\xi_0$ (see equation \ref{poly_radius})  
as a function of the normalized Yukawa length scale, for different values of the polytropic index $n$,
obtained from the numerical solution of equation (\ref{iter_equation}). The quantity along the x-axis actually is
 equal to $1/m\xi_0$, where $m^2=-1/3f''(0)$.}
\label{zmax_vs_lyuk} 
\end{figure}

\begin{figure}
\hspace{-0.4cm}
\centerline{\includegraphics[width=6.5cm, angle=-90]{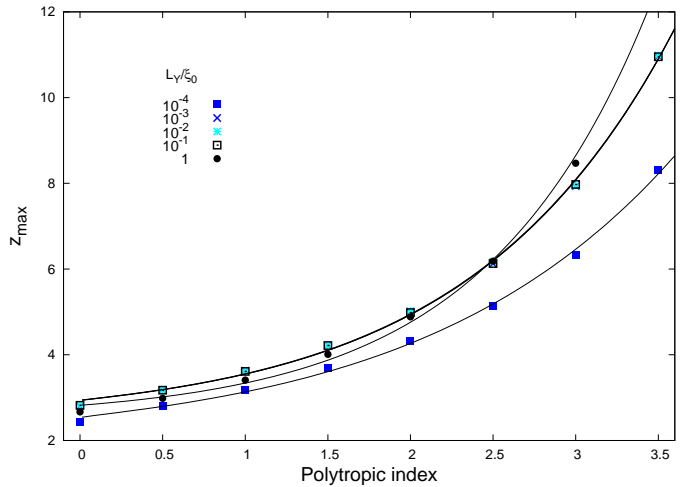}}
\caption{Stellar radius $\zmax$ in units of $\xi_0$ as a function of the polytropic index $n$ 
for different values of the normalized Yukawa length-scale $L_{\rm Y}/\xi_0$.} 
\label{zmax_vs_index} 
\end{figure}

As we are interested in finding self-consistent solutions, which in turn
correspond to determine the stellar radius of the polytropic star, 
it is worth pointing that $\zmax$  is not given by hand as a free parameter, but at each iteration 
it is determined by the value of $z$ such that $w(z)=0$.

In this way, the two input parameters required to derive a full solution
of equaton (\ref{le_equation}) are the  polytropic index $n$ and
the product $m~\xi_0$.
Bearing in mind that $m^2=-1/3f''(0)$ and $1/m=L_{\rm Y}$ defines the characteristic
scale of the Yukawa-like correction to the Newtonian potential, the quantity $1/m\xi_0=L_{\rm Y}/\xi_0$ just
expresses the Yukawa lenght scale in units of the polytropic radius $\xi_0$.
The fastness in the convergence of the iteration procedure depends on the initial guess 
of the $w(z)$ function, and we found that the setting $w^0(z)=1$ provides convergence after
just a few iterations.

\section{Results}
\label{results}

The robustness of the solving algorithm described in Section \ref{sect_algorithm} allowed
us the possibility to explore the solutions of the MLEE for any value of 
the poliyropic index $n < 5$ and the Yukawa length scale $L_{\rm Y}/\xi_0$ and
compare them with the case of GR.  The latter case simply corresponds to setting $m \xi_0=0$ in equation (\ref{le_equation}).

In Fig. \ref{wfunct_solutions} we report the function $w(z)$ for $ L_{\rm Y}/\xi_0=1$ and
four different peculiar values of the polytropic index $n$. 
For each value of $n$ we also reported the solution of the classical LE equation ($m \xi_0=0$);
we remind the reader the when $n=0$ and $n=1$ the classical solution admits the analytical expressions
  $w^0(z)=1-z^2/6$ and  $w^1(z)={\rm sin}(z)/z$, respectively.

The other two function $w(z)$ in Fig. \ref{wfunct_solutions} correspond to the polytropic
equation of state of a degenerate non-relativistic ($n=3/2$) and relativistic ($n=3)$ gas,
and can be of relevant astophysical interest for instance in the study stellar structure of compact objets.
As already known, for any fixed value of $n$, the radius is higher for $f(R)$-models than GR-models,
and the higher the polytropic index, the higher is the stellar radius $z_{\rm max}=R/\xi_0$.

To investigate how more generally $\zmax$ depends on the Yukawa length-scale value, we subsequently 
performed a set of runs with a grid of values of $L_{\rm Y}/\xi_0$ spanning several order
of magnitude, with results reported in Fig. \ref{zmax_vs_lyuk}.
For clarity plotting purposes we show results  for values of $n$ from 1 to 3
with stepsize of 0.5.
From the figure it can be seen that the deviations from the GR-values occur in a range
$10^{-4} \la L_{\rm Y}/\xi_0 \la 1$, which does not approximatively depend
on the politropic index $n$.
The region characterizing the deviations between GR and $f(R)$ solutions has 
a sharp increase around $L_{\rm Y}/\xi_0 \sim 10^{-4}$, followed  by a plateau
and a subsequent fast transition to the classical solution at $L_{\rm Y}/\xi_0 \sim 1$.
For $n \geqslant 2$ the right hand part of the plateau also shows a
secondary superposed increase of $\zmax$ before the steep decay to the classical
solution and specularly in the region around  $ L_{\rm Y}/\xi_0 \sim 10^{-4}$ 
the value of $\zmax$ is slightly lower than its GR-solution.

It is also worth noting that in all cases $ L_{\rm Y}/\xi_0 < \zmax $, 
or $L_{\rm Y} < R_{*}$, thus the Yukawa correction term, no matter
which is its true value in the considered region, acts on a characteristic length
scale lower than the stellar radius. We consider this point 
of key importance in the framework of the determination
of the stellar structure.

In Fig. \ref{zmax_vs_index} we report the behaviour of the stellar radius as
a function of the politropic index $n$ between 0 and 4 and for $10^{-4} \leq L_{y}/\xi_0  \leq 10$,
which corresponds to the range of Yukawa length-scales where
deviations occur from the classical solutions (see Fig. \ref{zmax_vs_lyuk}).
The data can be well fitted by a function $\zmax(n)=a ~{\rm exp}(-b~n)+c$
with  $a$ in the range $\sim  0.16\div0.27$, $b\sim 1$ and c in the range 
$\sim 2.4\div2.9$.

\section{Stellar mass and gravitational energy binding}
\label{cinque}

The solution of the LEE  allows, for given values of the central
density, the polytropic index $n$ and the constant $K$ of the 
polytropic relation $p=K \rho^{1+1/n}$, to find the mass and the gravitational 
binding energy of a polytropic star.
In the case of classical solution of the equation, the values are

\begin{equation}
 M=4 \pi \rho_{\rm c} R^3 \left(-\frac{1}{z}\frac{dw}{dz}\right)_{z=\zmax},
\label{mass_classical}
\end{equation}

\noindent
and
\begin{equation}
 E_{\rm g}=-\frac{3}{5-n}\frac{G M^2}{R}.
\label{egrav_classical}
\end{equation}

In more general case of MLEE, $E_{\rm g}$ and $M$ can be written
as 

\begin{equation}
 M=\int^{R_{\rm max}}_0 \rho({\bf x}) d{\bf x},
\label{mass_general}
\end{equation}

\noindent
and

\begin{equation}
 E_{\rm g}=\int^{R_{\rm max}}_0 \Phi({\bf x}) \rho({\bf x}) d{\bf x}.
\label{egrav_general}
\end{equation}

While keeping in mind the definitions $\Phi=w \Phi_{\rm c}$, $\Phi_{\rm c}= \rho_{\rm c}^{1/n} (n+1) K$,
and  $\rho=\rho_{\rm c} w^n$, equations (\ref{mass_general}) and (\ref{egrav_general}) can be rewritten
as 
\begin{equation}
 M= 4\pi \xi^3_0  \rho_{\rm c} \int^{z_{\rm max}}_0   w^{n} z^2 dz,
\label{mass_wz}
\end{equation}

and
\begin{equation}
 E_{\rm g}= -4\pi \xi^3_0 (n+1) K \rho^{1/n+1}_{\rm c} \int^{z_{\rm max}}_0   w^{n+1} z^2 dz,
\label{egrav_wz}
\end{equation}

\noindent
where $w(z)$ is a solution of the LE equation (\ref{le_equation}), and depends on
$n$ and $m~\xi_0$.
When $m~\xi_0 \rightarrow 0$, equations (\ref{mass_general}) and (\ref{egrav_general}) can
be solved analytically and reduce to equations (\ref{mass_classical}) and (\ref{egrav_classical}), respectively,
while in the general case of $f(R)$-gravity they need to be solved numerically.

Let us now consider for any given polytropic index $n$, the solutions of equation (\ref{le_equation}) for
the case of $f(R)$ and GR gravity, which we label as  $w_{\rm f(R)}$ and  $w_{\rm GR}$, respectively.
Using equations (\ref{mass_wz}) - (\ref{egrav_wz}) we can now write

\begin{equation}
 \frac{M_{\rm f(R)}}{M_{\rm GR}}= \frac{\int^{z_{\rm max}}_0   w_{f(R)}^{n} z^2 dz}{\int^{z_{\rm max}}_0   w_{GR}^{n}  z^2 dz}\,,
\label{ratio_integral_mass}
\end{equation}
\begin{equation}
 \frac{E^{\rm g}_{\rm f(R)}}{E^{\rm g}_{\rm GR}}= \frac{\int^{z_{\rm max}}_0   w_{f(R)}^{n+1} z^2 dz}{\int^{z_{\rm max}}_0   w_{GR}^{n+1}  z^2 dz}\,.
\label{ratio_integral_energy}
\end{equation}

\begin{figure}
\hspace{-0.4cm}
\centerline{\includegraphics[width=6.5cm, angle=-90]{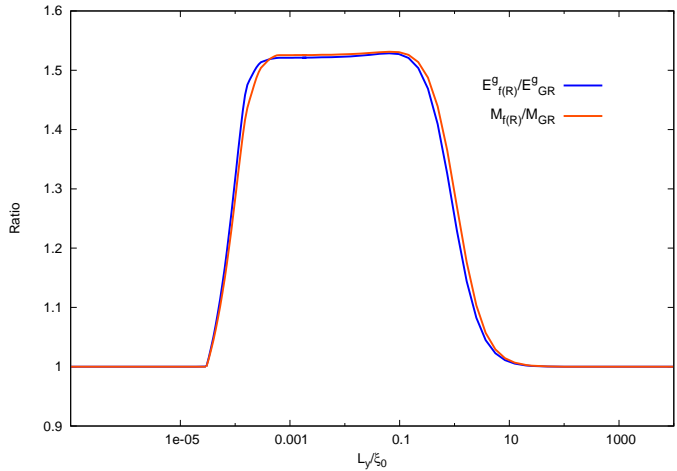}}
\caption{Ratio of the masses and  gravitational  binding  energies for $f(R)$ and GR polytropic stars 
(see eqs. \ref{ratio_integral_mass}-\ref{ratio_integral_energy} as a function of the
normalized Yukawa lenght scale $L_{\rm Y}/\xi_0$.  The results here reported are obtained for
a politropic index value $n=3$ but actually they are the same for any value of $n$ (see text).}
\label{ratio_mass_egrav_vs_lyuk} 
\end{figure}

It is evident that equations (\ref{ratio_integral_mass}) and  (\ref{ratio_integral_energy}) are very useful for
finding relative masses and gravitational  binding energies between $f(R)$ and GR stellar models, for any 
desired choice of the parameters $K$ and $\rho_c$, which indeed do not appear. 
The results of these ratios are reported in Fig. (\ref{ratio_mass_egrav_vs_lyuk}) for a range of values of $L_{\rm Y}/\xi_0$

Looking at Fig. (\ref{ratio_mass_egrav_vs_lyuk}), it is worth noting
that the value of the ratios  $M_{\rm f(R)}/M_{\rm GR}$ and  $E_{\rm f(R)}/E_{\rm GR}$ are very closed each other.
This is not surprising, given that in equations (\ref{ratio_integral_mass}) and (\ref{ratio_integral_energy}) 
the only difference is the change in the power of the $w(z)$-function from $n$ to $n+1$.

The most relevant result is that in $f(R)$-gravity masses and associated gravitational energy
binding values can exceed up to 50\%  of their classical value for $10^{-3} \la L_{\rm Y}/\xi_0 \la 10^{-1}$. 
Furthermore, we have two  characteristic gravitational radii, the standard Schwarzschild one
\begin{equation}
 R_{\rm s}=\frac{2 G M}{ c^2}\,,
\end{equation}
where $M$ is defined by equation (\ref{mass_wz}) and, as we have seen,  a length related to the Yukawa correction.
Keeping in mind the definition of $\xi_0$, equation  (\ref{poly_radius}),  and 
defining $L_{\rm Y}= \alpha \xi_0 $, where $\alpha= 1/m \xi_0$ and
$m^2=-1/3 f''(0)$, we obtain

\begin{equation}
\frac{ L_{\rm Y}}{ R_{\rm s}}= \frac{\alpha 2 c^2 }{ 3 K \rho_c^{1/n} (1+n) F(w,n)}\,,\label{ratio_lyuk_rscw}
\end{equation}
where 
\begin{equation}
 F(w,n)=\int^{\zmax}_0 w^n z^2 dz\,.
\end{equation}

For values of $n$ in the range $1\div4$, the ratio defined in equation (\ref{ratio_lyuk_rscw}) ranges from  $\sim \alpha\,10^4 K^{-1}_{14}\left( \rho_{100}\right)^{-1}$ to  $\sim \alpha\, 10^5 K^{-1}_{14}\left( \rho_{100}\right)^{-\frac{1}{4}}$ where $\rho_{100}=10^{-2}\rho_c$ and $K_{14}=10^{-14} K$. Looking at  Fig. \ref{ratio_mass_egrav_vs_lyuk}, the range of values of $\alpha=L_{\rm Y}/\xi_0$ for which significant deviations from standard Newtonian theory occur is $10^{-4} \la \alpha \la 0.1 $. This means that a second  gravitational radius, acting on scales larger than  the Schwarzschild radius, can be defined. From a dynamical point of view,  in principle,  this fact can be interpreted as the possibility to stabilize more massive self-gravitating systems that cannot be achieved in standard GR. In other words, very massive exotic objects that could not be obtained as solutions of the standard theory of stellar structures could be achieved in the framework of $f(R)$-gravity.

\section{Discussions and conclusions}
\label{sei}

%
%
%
%
%

The developement of an efficient algorithm aimed to solve the Lane-Emden equation for stellar structures in $f(R)$-gravity
 allows to open unexplored possibilities  for self-graviting systems and stellar structures.
In particular, we have seen that higher-order gravitational corrections   give the possibility to extend the range of stable stellar structures  by inducing a  further characteristic gravitational radius that, if compared with the Schwarzschild one can rule these new structures.

In general, one can get more massive structures that remain stable. These toy-models have of course to be refined considering also 
  values of physical interest  for  the costant $K$ of the polytropic relation, the central density $\rho_c$ and other secondary parameters. 
Specifically, a further numerical step and refinement will  consist in building up realistic  stellar models where the structure is 
divided into a given number of shells, each  with its own polytropic equation $p=K_i \rho^{\gamma_i}$.
For each $i^{\rm th}$ shell, one  has to  solve equation (\ref{le_equation}) by using,
as Cauchy initial conditions, the values $w(z^{i-1}_{\rm out})$ and  $w'(z^{i-1}_{\rm out})$ where
$z^{i-1}_{\rm out}$ corresponds to the outer border of the bottom $({i-1})^{\rm th}$ shell.
This would be particularly desired for main sequence model with a central nuclear burning core surrounded
by a convective envelope. Besides, the algorithm could work also to describe structures out of the main sequence,  peculiar objects and systems in the instability strip since boundary conditions on the MLEE are less restrictive with respect to those requested for standard LEE.
Finally, as discussed in \cite{Astashenok}, in the case of observation of self-gravitating systems that escape standard description of GR, this could constitute a straightforward testbed  for alternative theories of gravity.

\section*{Acknowledgments}

MDL is supported by MIUR (PRIN 2009). MDL and SC acknowledge the
support of INFN Sez. di Napoli (Iniziativa Specifica TEONGRAV).
%


\end{document}